\documentclass[a4paper,11pt]{article}
\usepackage{jinstpub} % for details on the use of the package, please see the JINST-author-manual
\usepackage{lineno}
\usepackage{adjustbox}

\usepackage{siunitx}
\newcommand{\mum}{\mathrm{\mu m}}

\newcommand{\frep}{f_{\mathrm{rep}}}
\newcommand{\nb}{n_{\mathrm{b}}}

\title{Beam Delivery and Beamstrahlung Considerations for Ultra-High Energy Linear Colliders}

\author[a]{Tim Barklow,}
\author[a]{Spencer Gessner,}
\author[a]{Mark Hogan,}
\author[a]{Cho-Kuen Ng,}
\author[a]{Michael Peskin,}
\author[a]{Tor Raubenheimer,}
\author[a]{Glen White,}

\affiliation[a]{SLAC National Accelerator Laboratory, Menlo Park, California, USA}

\author[b]{Erik Adli,}
\author[b]{Gevy Jiawei Cao,}
\author[b]{Carl A. Lindstr{\o}m,}
\author[b]{Kyrre Sjobak,}
\affiliation[b]{Department of Physics, University of Oslo, Oslo, Norway}

\author[c]{Sam Barber,}
\author[c]{Cameron Geddes,}
\author[c]{Arianna Formenti,}
\author[c]{Remi Lehe,}
\author[c]{Carl Schroeder,}
\author[c]{Davide Terzani,}
\author[c]{Jeroen van Tilborg,}
\author[c]{Jean-Luc Vay,}
\author[c]{Edoardo Zoni,}
\affiliation[c]{Lawrence Berkeley National Laboratory, Berkeley, California, USA}

\author[d]{Christopher Doss,}
\author[d]{Michael Litos,}
\affiliation[d]{University of Colorado Boulder, Boulder, Colorado, USA}

\author[e]{Ihar Lobach,}
\author[e]{John Power,}
\affiliation[e]{Argonne National Laboratory, Lemont, Illinois, USA}

\author[f]{Maximilian Swiatlowski}
\affiliation[f]{TRIUMF, Vancouver, Canada}

\author[g]{Luca Fedeli,}
\author[g]{Henri Vincenti,}
\affiliation[g]{Universite Paris-Saclay, CEA, CNRS, LIDYL, 91191 Gif-sur-Yvette, France}

\author[h]{Thomas Grismayer,}
\author[h]{Marija Vranic,}
\author[h]{Wenlong Zhang,}
\affiliation[h]{GoLP/IPFN, Instituto Superior Técnico, University of Lisbon, Lisbon, Portugal}

\emailAdd{sgess@slac.stanford.edu}

\abstract{As part of the Snowmass'21 community planning excercise, the Advanced Accelerator Concepts (AAC) community proposed future linear colliders with center-of-mass energies up to 15 TeV and luminosities up to 50$\times10^{34}$ cm$^{-2}$s$^{-1}$ in a compact footprint. In addition to being compact, these machines must also be energy efficient. We identify two challenges that must be addressed in the design of these machines. First, the Beam Delivery System (BDS) must not add significant length to the accelerator complex. Second, beam parameters must be chosen to mitigate beamstrahlung effects and maximize the luminosity-per-power of the machine. In this paper, we review advances in plasma lens technology that will help to reduce the length of the BDS system and we detail new Particle-in-Cell simulation studies that will provide insight into beamstrahlung mitigation techniques. We apply our analysis to both $e^+e^-$ and $\gamma\gamma$ colliders.}

\begin{document}
\maketitle
\flushbottom

\section{Introduction}
\label{sec:intro}

Research on Advanced Accelerators Concepts (AAC) for future Linear Colliders has, to this point, been focused on creating high-gradient, high-efficiency, high-quality plasma and structure wake-field accelerators. Just as critical to the performance of a future Advanced Linear Collider is the Beam Delivery System (BDS), which delivers beam to the interaction point~\cite{White2022}. The BDS must be compact in order to minimize the footprint of the accelerator facility. In addition the beam parameters must be chosen so as to minimize beamstrahlung effects and maximize the luminosity of the machine for a given input power.

\begin{figure}
    \centering

    \includegraphics[width=0.8\linewidth]{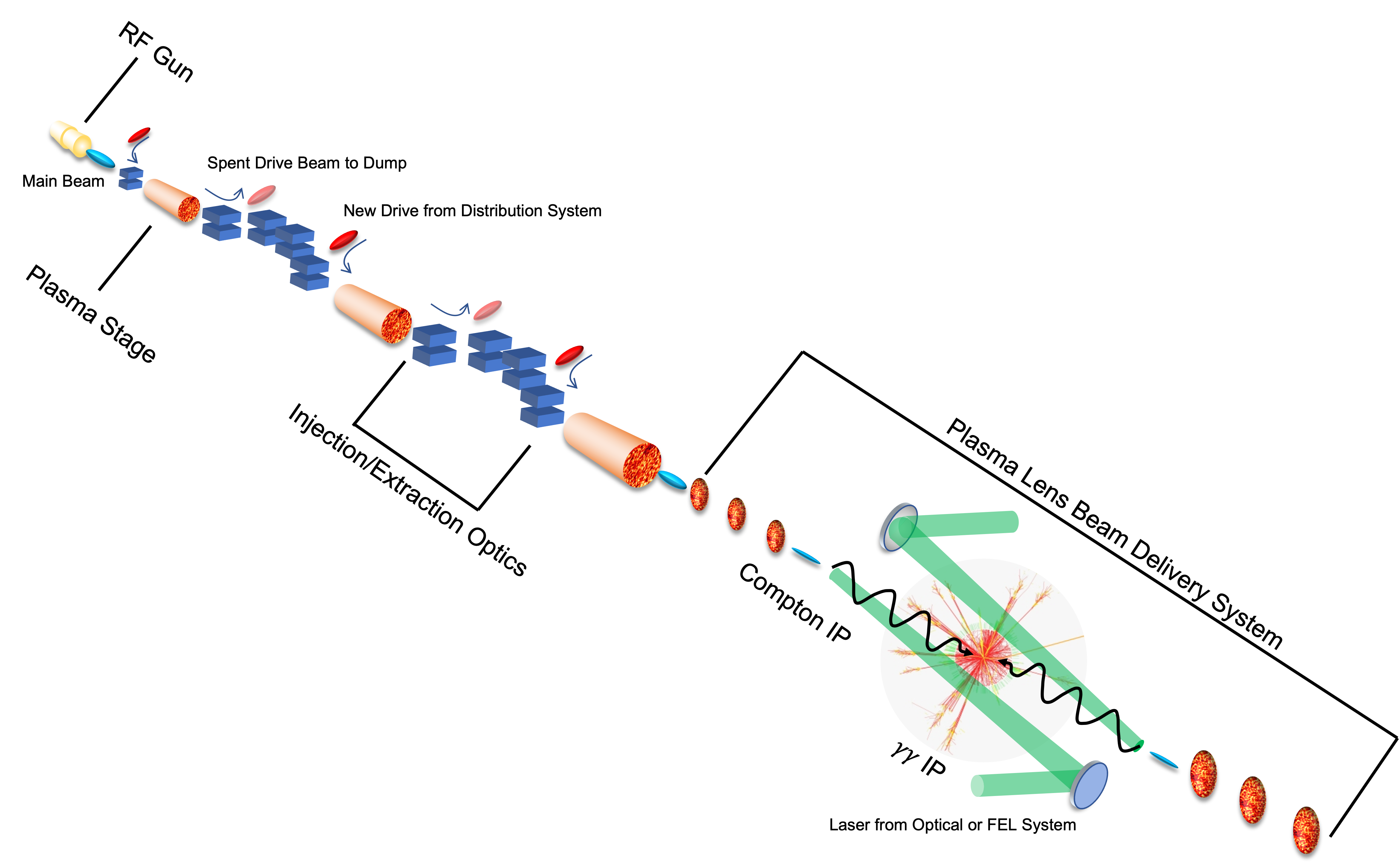}
    \caption{Illustration of a $\gamma-\gamma$ linear collider with plasma lens focusing elements.}
    \label{fig:gamgam}
\end{figure}

Figure~\ref{fig:gamgam} shows a concept for a $\gamma-\gamma$ linear collider with plasma lens focusing elements. This type of collider will require a completely re-imagined BDS. The BDS system must maximize luminosity, while respecting the constraints of the particle detector, incoming and outgoing particle beams, and incoming and outgoing laser beams.

In this paper we summarize the challenges associated with creating a compact BDS system and explore alternative designs for BDS systems, including plasma lenses. We detail the challenges of operating a collider at large $\Upsilon$ (high beamstrahlung) and detail new efforts to better simulate and understand beamstrahlung with Particle-in-Cell (PIC) codes. Finally, we discuss the Machine-Detector Interface (MDI) in the presence of plasma and speculate on novel detector designs.

\section{Discussion of the Snowmass Parameter sets for Advanced Linear Colliders}

For Snowmass'21, the Advanced Accelerator Concepts (AAC) community proposed future linear colliders with center-of-mass energies up to 15 TeV and luminosities up to 50$\times10^{34}$ cm$^{-2}$s$^{-1}$. These values can be achieved by suitable choices of beam parameters, but the power consumption of the machine must also be taken into consideration. In order to minimize power consumption while maximizing the luminosity of the machine, a number of new concepts were explored, including round-beam collisions. Table~\ref{ref:wfa_table} summarizes the parameters explored for different accelerator technologies at different beam energies and spot-size ratios, as documented in the ITF report~\cite{ITFReport}.

\begin{table}\label{ref:wfa_table}

\small 
    \centering
    \begin{adjustbox}{angle=90}
    \begin{tabular}{|l|l|l|l|l||l|l|l||l|l|l|}
    \hline
        Technology & PWFA & PWFA & PWFA & PWFA & SWFA & SWFA & SWFA & LWFA & LWFA & LWFA \\ \hline
        Beam Aspect Ratio & Flat & Flat & Flat & Round & Flat & Flat & Round & Flat & Flat & Round \\ %\hline
        Center-of-Mass Energy & 1 & 3 & 15 & 15 & 1 & 3 & 15 & 1 & 3 & 15 \\ %\hline
        $E_{beam}$ (TeV) & 0.5 & 1.5 & 7.5 & 7.5 & 0.5 & 1.5 & 7.5 & 0.5 & 1.5 & 7.5 \\ %\hline
        $\gamma$ & 9.78E5 & 2.94E6 & 1.47E7 & 1.47E7 & 9.78E5 & 2.94E6 & 1.47E7 & 9.78E5 & 2.94E6 & 1.47E7 \\ %\hline
        $\varepsilon_x$ (mm mrad) & 0.66 & 0.66 & 0.66 & 0.1 & 0.66 & 0.66 & 0.1 & 0.1 & 0.02 & 0.1 \\ %\hline
        $\varepsilon_y$ (mm mrad) & 0.02 & 0.02 & 0.02 & 0.1 & 0.02 & 0.02 & 0.1 & 0.01 & 0.007 & 0.1 \\ %\hline
        $\beta^*_x$ (mm) & 5 & 5 & 5 &  0.15 & 5 & 5 & 0.15 & 25 & 14 & 0.15 \\ %\hline
        $\beta^*_y$ (mm) & 0.1 & 0.1 & 0.1 & 0.15 & 0.1 & 0.1 & 0.15 & 0.1 & 0.1 & 0.15 \\ %\hline
        $\sigma^*_x$ (nm) & 58.07 & 33.53 & 15 & 1.01 & 58.07 & 33.53 & 1.01 & 50.55 & 9.77 & 1.01 \\ %\hline
        $\sigma^*_y$ (nm) & 1.43 & 0.83 & 0.4 & 1.01 & 1.43 & 0.83 & 1.01 & 1.01 & 0.49 & 1.01 \\ %\hline
        $N_{bunch}$ ($\times 10^9$) & 5 & 5 & 5 & 5 & 3.13 & 3.13 & 3.13 & 1.20 & 1.20 & 7.50 \\ %\hline
        $f$ (kHz) & 4.2 & 14 & 13.12 & 7.73 & 11 & 36 & 19.8 & 46.8 & 46.8 & 3.44 \\ %\hline
        $\sigma_z$ (um) & 5 & 5 & 5 & 5 & 40 & 40 & 40 & 8.4 & 8.4 & 2.2 \\ %\hline
%        ~ & ~ & ~ & ~ & ~ & ~ & ~ & ~ & ~ & ~ \\ %\hline
        $\Upsilon$ & 15 & 78 & 867 & 6590 & 1 & 6 & 515 & 2 & 37 & 22466 \\ %\hline
        $n_{\gamma}$ & 1.5 & 1.5 & 1.5 & 5.7 & 2.2 & 2.2 & 8.4 & 0.8 & 1.5 & 5.7 \\ \hline
%        ~ & ~ & ~ & ~ & ~ & ~ & ~ & ~ & ~ & ~ \\ \hline
        $P_{beam}$ (MW) & 1.7 & 16.8 & 78.8 & 55.0 & 2.8 & 27.0 & 74.4 & 4.5 & 13.5 & 31.0 \\ %\hline
        $2P_{beam}$ (MW) & 3.4 & 33.6 & 157.6 & 110.0 & 5.5 & 54.1 & 148.7 & 9.0 & 27.0 & 61.9 \\ \hline
%        ~ & ~ & ~ & ~ & ~ & ~ & ~ & ~ & ~ & ~ \\ %\hline
        $\mathcal{L}_{geo}$ ($\times 10^{34}$ cm$^{-2}$s$^{-1}$) & 1.01 & 10.1 & 47.1 & 150 & 1.03 & 10.1 & 151 & 1.05 & 11.3 & 150 \\ %\hline
        $\mathcal{L}_{beamstrahlung}$ ($\times 10^{34}$ cm$^{-2}$s$^{-1}$) & 1.99 & 19.9 & 99.4 & 152 & 2.03 & 20 & 152 & 2.09 & 21.7 & 152 \\ %\hline
%        ~ & ~ & ~ & ~ & ~ & ~ & ~ & ~ & ~ & ~ \\ \hline
        $\eta_{wall-to-drive}$ & 0.4 & 0.4 & 0.4 & 0.4 & 0.774 & 0.774 & 0.774 & 0.4 & 0.4 & 0.5 \\ %\hline
        $\eta_{drive-to-main}$ & 0.375 & 0.375 & 0.375 & 0.375 & 0.42 & 0.42 & 0.42 & 0.2 & 0.2 & 0.12 \\ %\hline
        $\eta_{total}$ & 0.15 & 0.15 & 0.15 & 0.15 & 0.325 & 0.325 & 0.325 & 0.08 & 0.08 & 0.06 \\ %\hline
        $P_{site}$ (MW) & 22 & 224 & 1051 & 619 & 17 & 166 & 457 & 113 & 338 & 1032 \\ %\hline
        $\mathcal{L}_{geo}/P_{site}$ (1e34/MW) & 0.04 & 0.04 & 0.04 & 0.08 & 0.06 & 0.06 & 0.11 & 0.01 & 0.03 & 0.05 \\ \hline
%        ~ & ~ & ~ & ~ & ~ & ~ & ~ & ~ & ~ & ~ \\ \hline
        $\mathcal{L}_{GP,tot}$ ($\times 10^{34}$ cm$^{-2}$s$^{-1}$) & 1.83 & 18.5 & 87.6 & 1570 & 2.08 & 21.3 & 420 & 1.53 & 25.8 & 600 \\ %\hline
        $\mathcal{L}_{GP,top}$ 1\% (20\%) ($\times 10^{34}$ cm$^{-2}$s$^{-1}$) & 0.69 & 6.23 & 50 & 50 & 0.85 & 6.14 & 50 & 1.03 & 8.72 & 50 \\ %\hline
        $\mathcal{L}_{GP,tot}/P_{site}$ & 0.08 & 0.08 & 0.08 & 2.54 & 0.12 & 0.13 & 0.92 & 0.01 & 0.08 & 0.58 \\ %\hline
        $\mathcal{L}_{GP,top}/P_{site}$ 1\% (20\%)/Power  & 0.03 & 0.03 & 0.05 & 0.08 & 0.05 & 0.04 & 0.11 & 0.01 & 0.03 & 0.05 \\ \hline
 %       ~ & ~ & ~ & ~ & ~ & ~ & ~ & ~ & ~ & ~ \\ \hline
        Length of 2 Linacs (km) & 1 & 3 & 14 & 14 & 5 & 15 & 75 & 0.44 & 1.3 & 6.5 \\
%        \hline
        Length of Facility & 14 & 14 & 14 & 14 & 8 & 18 & 90 & 3.5 & 4.5 & 9.5 \\ \hline
    \end{tabular}
    \end{adjustbox}
\caption{Parameters of the advanced WFA-based colliders. $\Upsilon$ is the beamstrahlung parameter. $\mathcal{L}_{GP,tot}$ and $\mathcal{L}_{GP,top}$ refer to the total and top percentage luminosity derived from GUINEA-PIG simulations. At 1 and 3 TeV CM, the top percentage luminosity is within 1\% of the CM energy. At 15 TeV CM, the top percentage luminosity is within 20\% of the CM energy. The simulations shown in Figure~\ref{fig:compFlatRound} are represented by the data in the ``PWFA Flat, 15 TeV" and ``PWFA Round, 15 TeV" columns.}

\end{table}

\begin{figure}[tb]
\centering
\includegraphics[width=5in, keepaspectratio]{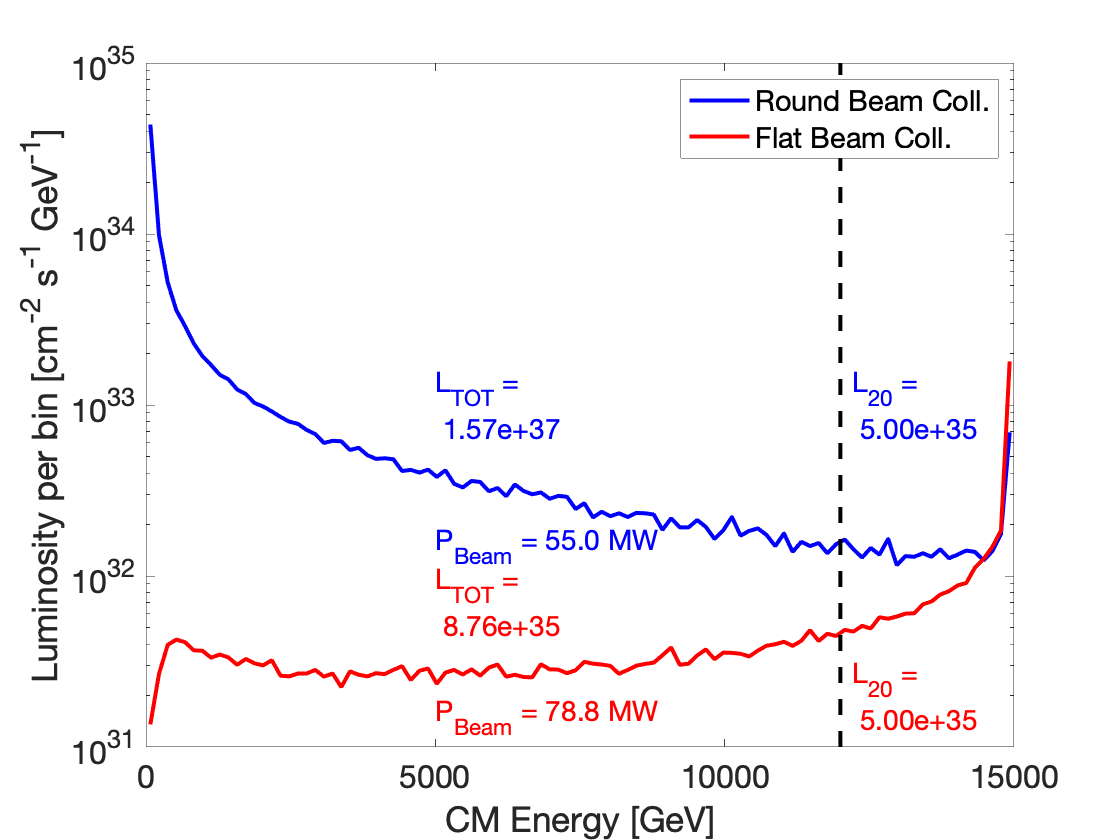}
\caption[]{Comparison of luminosity spectra for round beams and flat beams at 15 TeV Center-of-Mass energy.}
\label{fig:compFlatRound}
\end{figure}

The exploration of alternative spot-size ratios was illuminating. In particular, it revealed that although round beam collisions do increase the total luminosity for a given beam power, much of that luminosity increase comes from particle collisions at low energy, as shown in Figure~\ref{fig:compFlatRound}. This is due to particle energy loss from beamstrahlung. The luminosity of a linear collider is often characterized by the total luminosity $\mathcal{L}$ and the luminosity within 1\% of the desired center-of-mass energy $\mathcal{L}_{0.01}$. Linear colliders are often considered to be precision machines, and therefore $\mathcal{L}_{0.01}/\mathcal{L}$ should be as close to 1 as possible. For energy frontier colliders exploring generalized extensions to the Standard Model, such as a linear collider operating at 15 TeV CM, the value of $\mathcal{L}_{0.01}$ is less important, because there is no particular resonance or energy threshhold that must be achieved. We adopted a new figure-of-merit for these machines: luminosity within 20\% of the CM energy $\mathcal{L}_{0.20}$. This definition allows for a more direct comparison with other energy frontier machines such as FCC-hh.

\subsection{$\gamma-\gamma$ Colliders}
\begin{figure}[tb]
\centering
\includegraphics[width=5in, keepaspectratio]{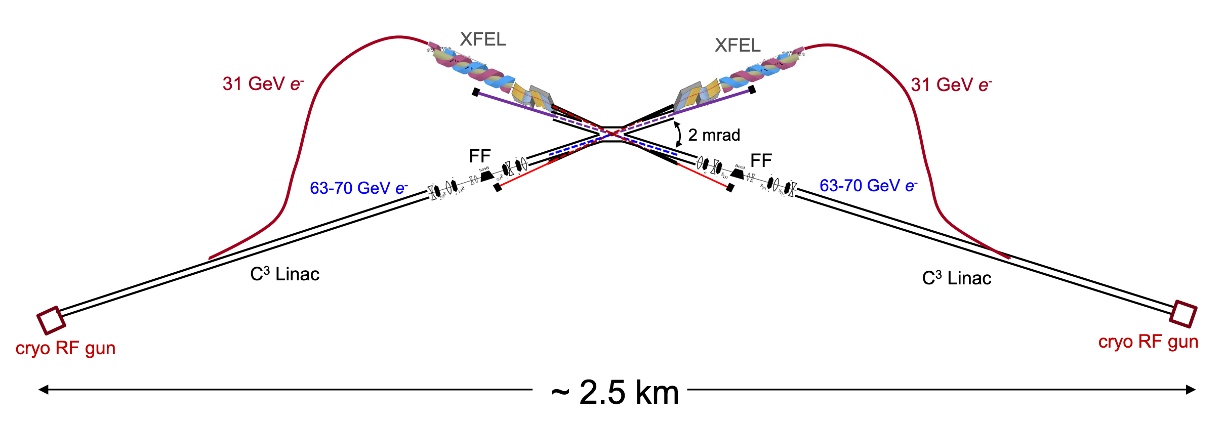}
\caption[]{The XCC Higgs Factory reproduced from Reference~\cite{BarklowXcc}.}
\label{fig:xcc}
\end{figure}

A $\gamma-\gamma$ Higgs factory at 125 GeV CM was proposed as part of the Snowmass'21, with the novelty of using x-ray laser beams for the Compton scattering source~\cite{BarklowXcc}. Figure~\ref{fig:xcc} illustrates the design of the XCC collider.

\begin{figure}[tb]
\centering
\includegraphics[width=5in, keepaspectratio]{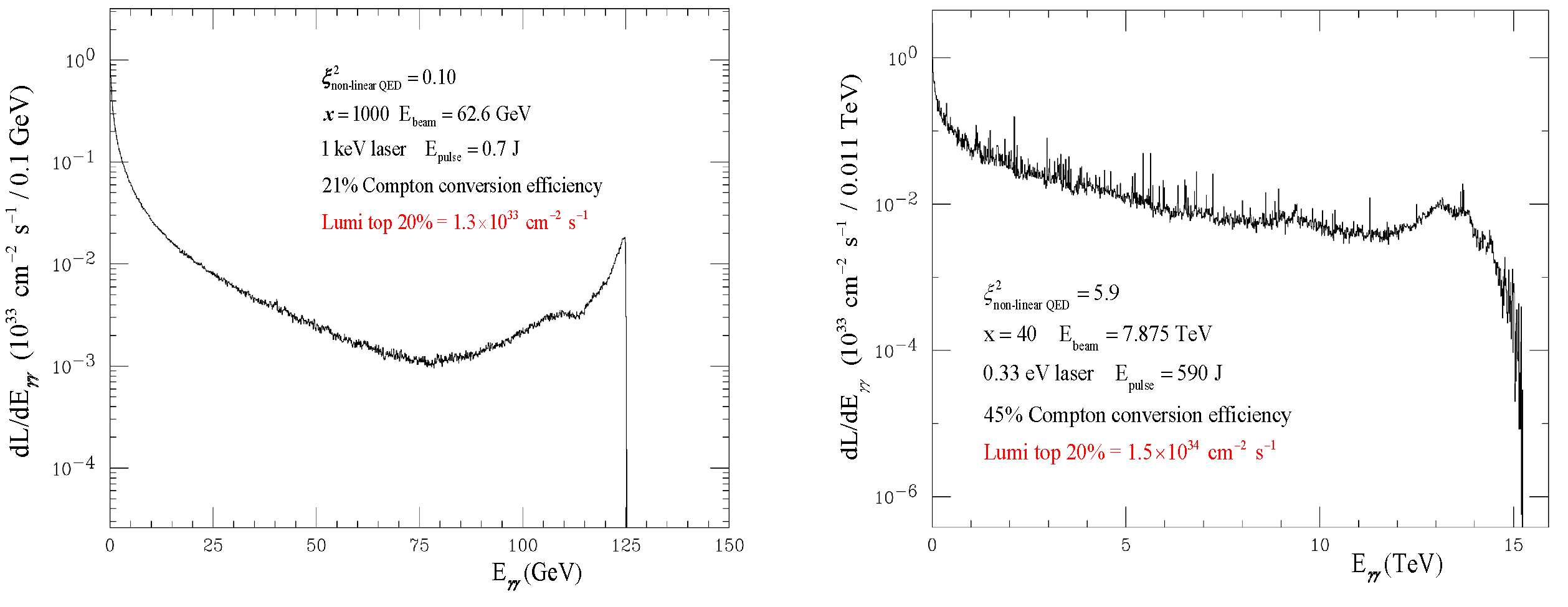}
\caption[]{$\gamma-\gamma$ collisions using the CAIN code. On the left, we see that for the Higgs Factory parameters, a relatively clean luminosity spectrum is produced. On the right, we observe a more complicated luminosity spectrum due to non-linear effects.}
\label{fig:cain}
\end{figure}

The CAIN code~\cite{Chen:1994jt} was used to simulate the Compton scattering and $\gamma-\gamma$ collision. Figure~\ref{fig:cain} shows that it is possible to simulate a relative clean spectrum in the case of $\gamma-\gamma$ collisions at 125 GeV CM, but non-linear effects are important at 15 TeV CM and they dilute the spectrum. The parameters for 125~GeV have been optimized to maximize the production of Higgs bosons (spike at 125 GeV) while minimizing background (low top 20\% luminosity).  The exploration of the parameter space for 15~TeV CM has just started.
% a little more discussion about x= 4.8 vs large x regime
% ran sims to get luminosity spectra for round beams, but have not compared to flat
% round beams non-optimal because left-over electron beams still tightly
% IP design for gamma-gamma?
% NLC ZDR, TESLA paper, CLICHE gamma-gamma

\section{\label{sec:escaling}Energy Scaling of BDS Length}
% Scaling with beta^*, lower beta^* -> more chromaticity correction needed
% Scaling with round vs flat, compare geometric mean of the beta functions

For a plasma-based linear collider, the length of the BDS system might exceed the length of the linac, and at very high energies, the BDS might comprise most of the machine. We attempt here to examine how the overall length of the BDS changes with design beam energy. We use the baseline design of the ILC or CLIC BDS as a reference (see Table~\ref{tab:BDSLEN}). We examine the BDS in 3 different sections: the final focus system (FFS), the bending systems (various diagnostics and energy collimation chicanes), and the other systems (transport, matching, betatron collimation and other beam diagnostics sections).

\begin{table}[h!]
\centering
\begin{tabular}{||l l l||} 
 \hline
  & ILC ($E_{cm} = 1 TeV$) & CLIC ($E_{cm} = 3 TeV$) \\ [0.5ex] 
 \hline\hline
 FFS & 826 m & 446 m \\ 
 \hline
 Bending Sections & 562 m & 1250 m \\
 \hline
 Other & 866 m & 1054 m \\
 \hline\hline
 Total & 2254 m & 2750 m \\
 \hline
\end{tabular}
\caption{Path length of ILC and CLIC BDS.}
\label{tab:BDSLEN}
\end{table}

\subsection{Bending sections}
%Emittance degradation in the bending systems due to incoherent synchroton radiation is generated in the bending plane according to~\cite{ISREmit}
Emittance degradation in the bending systems due to incoherent synchroton radiation is generated in the bending plane according to~\cite{raubenheimer2000final}
\begin{equation}
    \Delta\gamma\epsilon \approx (4 \times 10^{-8} m^2 GeV^{-6}) E^6 \sum_i{\frac{L_i}{|\rho_i|^3}\mathcal{H}_i},
\end{equation} 
where $\rho$ is the bending radius and $\mathcal{H}$ is a function of the lattice parameters through the section. Making the assumption that these sections are dominated by the bending magnets and that the target lattice parameters and required target dispersion remain the same, we thus make the assumption that the length in these sections scale according to the need to increase the bend radius: $L \sim E^2$.

\subsection{Matching / transport / diagnostics / collimation sections}
The MPS and betatron collimation systems, coupling correction and emittance measurement systems, extraction systems, and matching optics between the various BDS subsystems are dominated by the quadrupole focusing magnets required to perform the desired betatron manipulations. We use here the energy scaling of a FODO system to represent the scaling behavior of this ensemble of systems. Using the fact that the BDS length for ILC has been optimized, including for overall length considerations, we scale the lattice length from the designed energy assuming a scaling of the quadrupole focusing length parameter with energy. Using the usual transport matrix approximation for a FODO cell with thin-lens quadrupole elements, it can be shown that the phase-advance is related to the length of the cell as:
$sin(\theta/2) = L_{cell} / (4*f)$.
Using a fixed phase advanced as a proxy to maintaining the functionality of the various subsystems, the length of these systems is scaled simply with $E$.

\begin{figure}
\centering
    \includegraphics[width=0.4\linewidth]{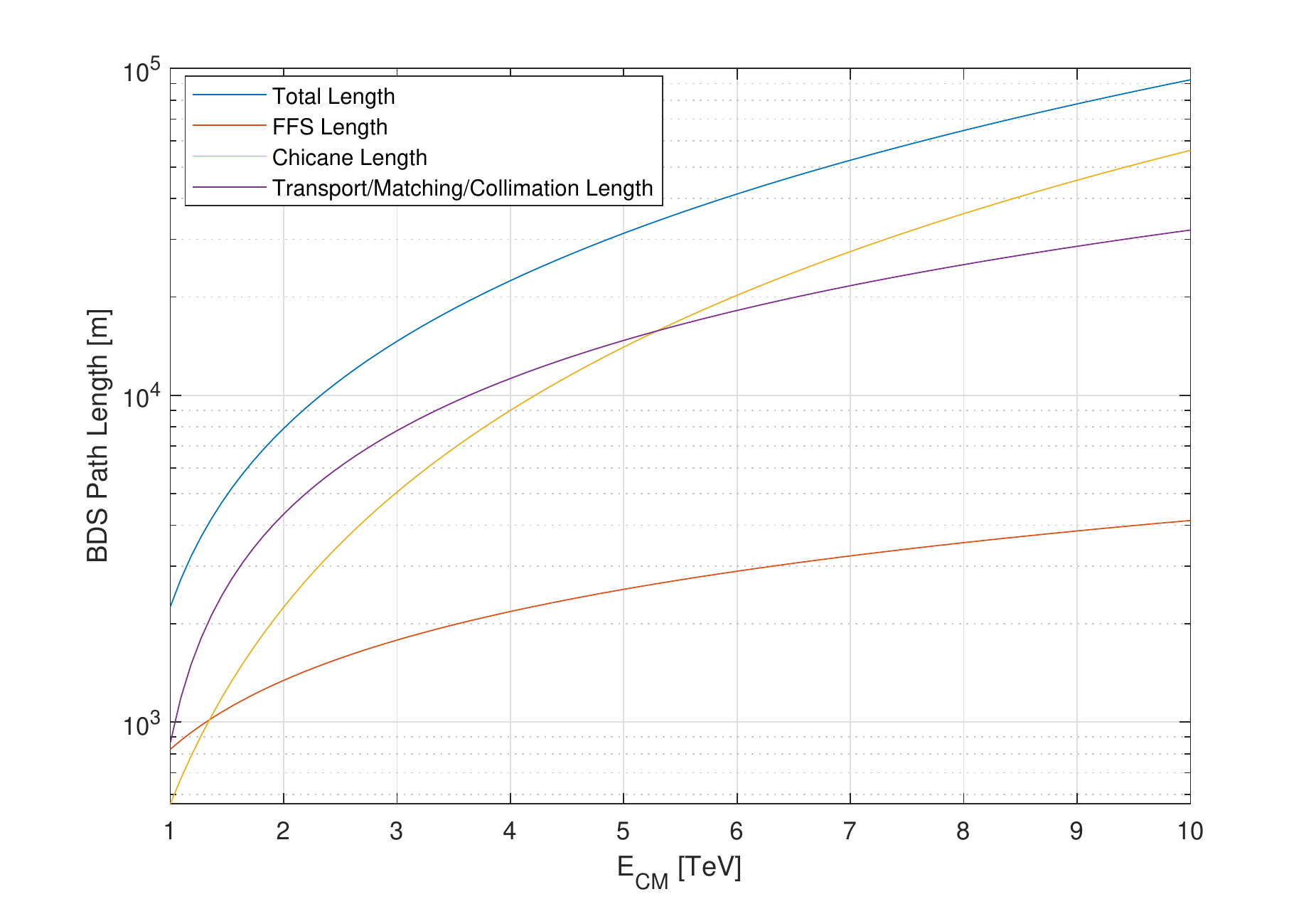} \includegraphics[width=0.4\linewidth]{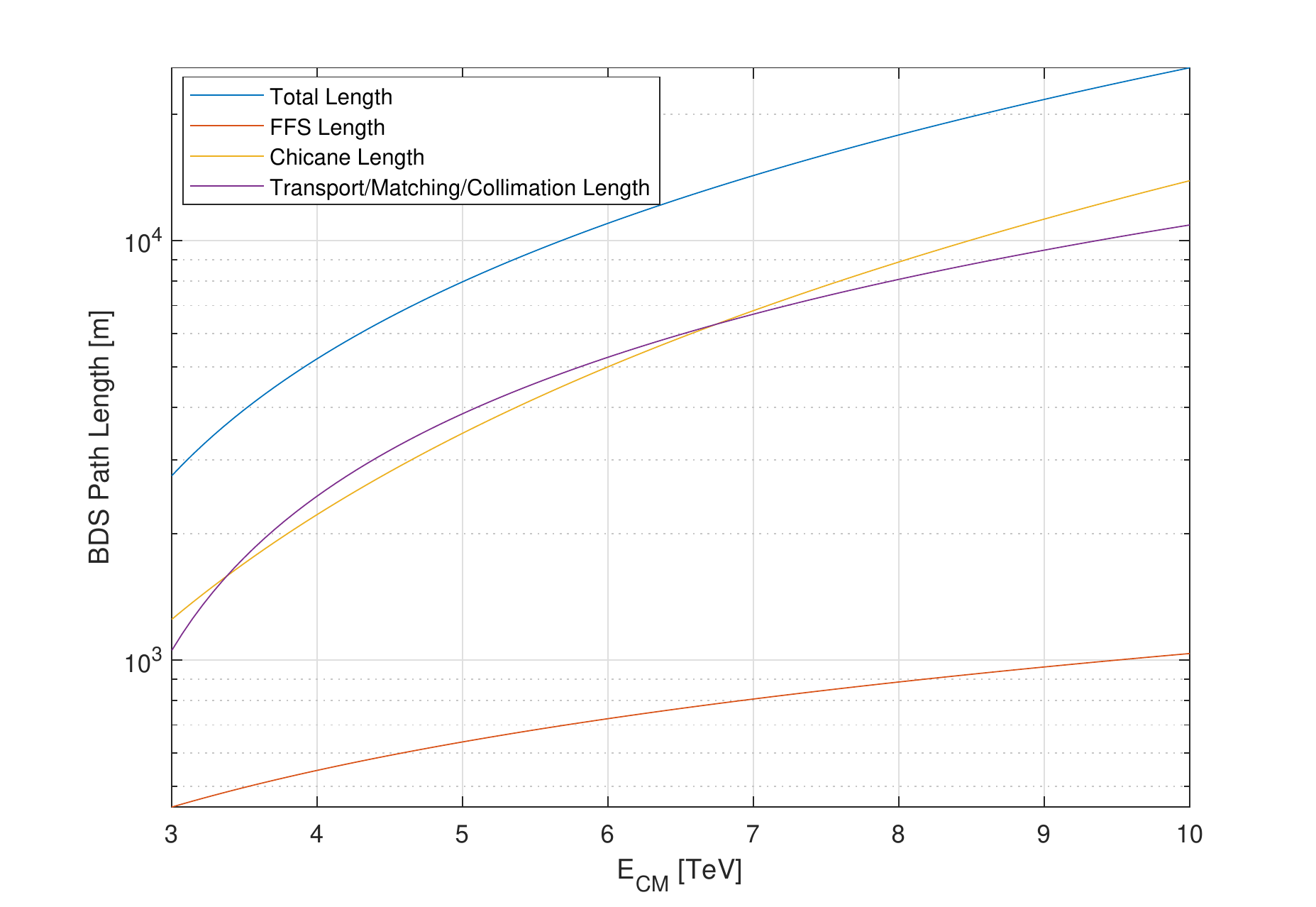}
    \caption{Path length of ILC (left) and CLIC (right) BDS as a function of collision energy using the scaling laws outlined in Section~\ref{sec:escaling}. "Chicane Length" refers to the sum of all bending sections.}
    \label{fig:CLICBDS}
\end{figure}

\subsection{Final focus system}
Scaling of the local-chromaticity correction style of final focus used in ILC is discussed in the original design paper for this style of optics by Raimondi/Seryi~\cite{Raimondi2001}. For fixed emittance, the FFS length scales as $E^{7/10}$, again the main constraint here is the emittance growth in the bending systems of the FFS. This does, however, make assumptions about the scalability of field gradients in the FFS quadrupoles, especially the final doublet, which may not be reasonable. Another assumption is that the IP beta functions and chromaticities remain constant (equivalent to holding $L^*$ and the length of the final doublet quadrupoles constant). This is also not reasonable due to limitations in possible quadrupole strengths and the need to achieve higher luminosities at higher energies. Also, there is freedom in the design of this style of FFS to trade-off length with tolerances (which relates to risk of operation through increased difficulty in tuning).

Given the complexities involved in the FFS, it is not really feasible to pin down an analytic scaling law that takes into account the changing needs of the system design as one moves to higher energies, or properly take into account the changing complexities of the tuning system without performing extensive design evaluations at each energy. For the sake of this analysis, we nevertheless use the only option of utilizing the $E^{7/10}$ scaling motivated by~\cite{Raimondi2001}.

\subsection{Overall scaling}

Whilst the final focus system itself looks to be scalable to the $\sim 10~TeV$ energy scale within a reasonable length footprint (few km) (with the many caveats stated earlier), the various other systems which comprise the BDS do not scale so favorably. Note however the discrepancy between the scaled length of ILC at 3 TeV and the custom design length of CLIC at that energy. This difference comes from the CLIC design being significantly more aggressive than ILC: more tolerant to risks and more optimistic about background conditions and ability to handle tolerances which are orders of magnitude tighter than ILC. Recent work has found that a BDS system for CLIC operating at 7 TeV CM appears feasible, and reducing the length of the collimation system has a modest effect on collider luminosity~\cite{Manosperti2023}.

\section{\label{sec:lenses}Plasma Lenses}
Plasma lenses are strong, axisymmetric focusing elements that may reduce the size of the BDS, while also reducing chromatic effects. In this section, we examine passive and active plasma lenses as possible elements of a final focus system.

\subsection{\label{sec:passivelens}Passive Plasma Lenses}

% General comment on alignment

Plasma lenses can focus electron beams with strengths several orders of magnitude stronger than quadrupole magnets~\cite{Doss2019,chen:1989prd,chen:1990prl}. In passive plasma lenses, the transverse focusing force in the underdense, nonlinear blowout plasma wake regime is due to the presence of the stationary plasma ions. If the transverse density profile of this ion column is uniform, then the focusing force experienced by the electrons in a relativistic beam is both axisymmetric and linear. These properties lead to an aberration-free focus of the beam that can achieve unprecedented small beam spots. The first order beam dynamics have been described in Ref.~\cite{Doss2019}.

\begin{figure}[tb]
\centering
\includegraphics[width=3.4in, keepaspectratio]{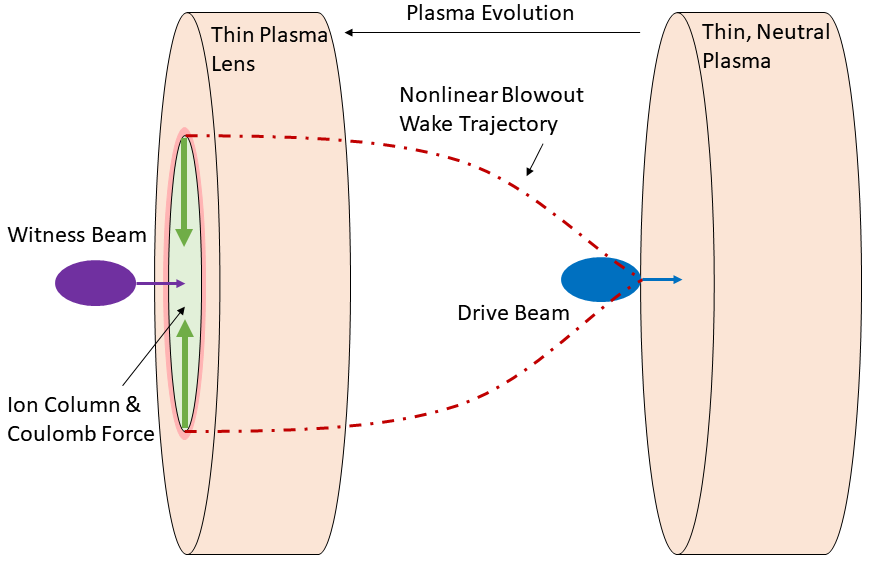}
\caption[]{Illustration of the underdense, passive plasma lens.  A dense drive beam drives a nonlinear blowout wake in an underdense neutral plasma, and a subsequent witness beam is focused by the Coulomb attraction of the stationary ions.  The force is axisymmetric and linear so long as the witness beam is fully within the blowout wake and the plasma density is uniform.}
\label{fig:gasjet}
\end{figure}

The underdense plasma lens (UPL) operates in a two-bunch configuration where a ``driver'' electron bunch drives the nonlinear wake and a second, ``witness'' electron bunch is subsequently focused in the nonlinear blowout wake (Fig.~\ref{fig:gasjet}).  It may also be possible to operate in a single-bunch configuration where the head of the electron bunch drives the nonlinear wake and the bulk of the electron bunch is focused.  In this regime, the linear focusing force from the ions results in a thin lens focal length given by the focusing strength
\begin{equation}\label{eqn:uplf}
f\equiv\frac{1}{KL}=\frac{1}{2 \pi r_e}\frac{\gamma_b}{n_p L}.
\end{equation}
Here, $K$ is the focusing strength, $L$ is the lens's longitudinal thickness, $r_e$ is the classical electron radius, $\gamma_b$ is the Lorentz factor of the electron beam, and $n_p$ is the plasma density of the lens.  A UPL with a plasma density of $n_p=1\times 10^{17}\mathrm{\ cm^{-3}}$ can have a focusing strength $K=88400\mathrm{\ m^{-2}}$ and focal length $f=3.3\mathrm{\ cm}$.  For comparison, a conventional quadrupole with field gradient $G=1\mathrm{\ T/m}$ would have $K=0.3\mathrm{\ m^{-2}}$ and focal length $f=1000\mathrm{\ cm}$ and a permanent magnetic quadrupole with $G=500\mathrm{\ T/m}$ can expect $K=150\mathrm{\ m^{-2}}$ and focal length $f=81\mathrm{\ cm}$.

The UPL can be generated by laser ionization of gas, e.g. by focusing a femtosecond laser pulse into a gas jet. To minimize the footprint, the laser pulse can propagate transverse to the electron beam axis, reducing the required space along the beam line to millimeters.  The longitudinal density profile of the plasma lens is then given by the laser parameters and focusing optics for gas ionization.  Something as simple as a Gaussian focus from a spherical lens with a variable offset from the laser focus to the beamline can generate a plasma lens with thickness in the range of 10-100 $\mu$m.

There has been discussion about the potential use of a plasma lens for the final focusing element of a collider for decades~\cite{chen:1989prd,chen:1990prl}. The strong, linear, axisymmetric focusing of the UPL in combination with its ultra-compactness and ``self-aligning'' characteristics may show it to be a viable candidate, although further study is still required. It should be noted that the functionality of the UPL described above applies only to negatively charged relativistic leptons. Relativistic positrons would experience a linear {\it defocusing} force in this scheme. It may be possible to achieve similarly strong focusing with the UPL if the positrons are sent through the lens in a different phase of the nonlinear wake, but this is speculative. Another important consideration is the scattering of the beam off of the plasma ions, which may produce both an increase in the beam emittance and a forward-directed shower of secondary particles. The former could result in a reduced luminosity and the latter could result in an increased background for the particle detectors.

\subsection{Active Plasma Lenses} 

\begin{figure}
    \centering
    %
    %\missingfigure[figwidth=\linewidth]{Tilborg figure which is used everywhere}
    %\caption{Illustration of the priniciple of an active plasma lens. From~[Tilborg et al., 2015].}
    %\todo[inline]{Replace with non-screenshoted variant. Too big? Set reference https://link.aps.org/doi/10.1103/PhysRevLett.115.184802}
    %
    \includegraphics[width=0.8\linewidth]{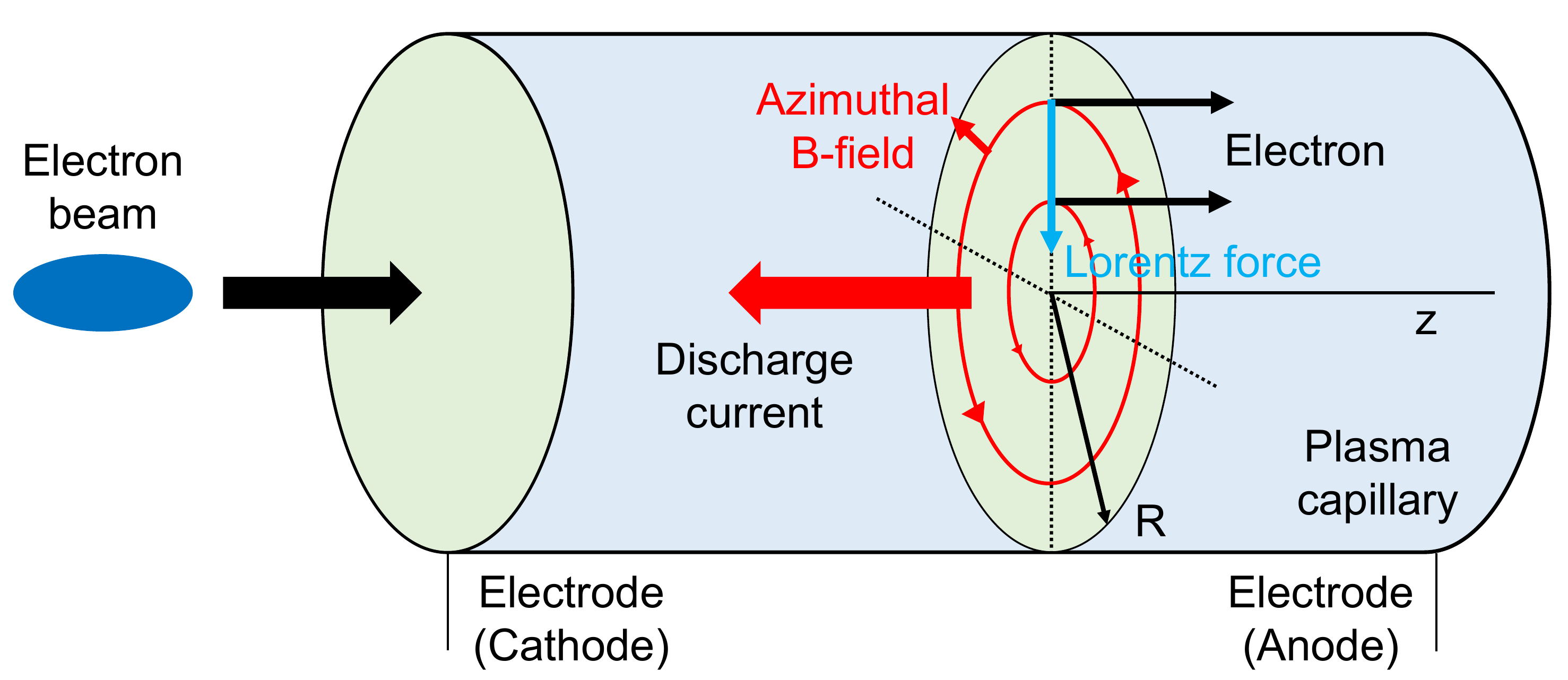}
    \caption{Schematic illustration of an active plasma lens. From~\cite{kim_witness_2021}.}
    \label{fig:APL:schematic}
\end{figure}

Active plasma lenses (APL), illustrated in Fig.~\ref{fig:APL:schematic}, are magnetic focusing devices that work by passing a strong current though a plasma, on the same axis as the beam.
This creates an azimuthal uniform magnetic field inside the plasma that produces  axisymmetric focusing, unlike a quadrupole that focuses in one plane and defocuses in the other. 
The field inside is given by the current distribution in the lens, and in the ideal case with a uniform current distribution and with a circular aperture with radius $R$, we have the magnetic field as a function of radial position $r$
\begin{equation}
    B_\phi(r) = \frac{\mu_0 I}{2\pi} \frac{r}{R^2} = g_\mathrm{APL} r\,.
\end{equation}
Here $I$ is the total current in the lens, and $g_\mathrm{APL}$ the magnetic gradient. Current APLs typically use a gas-filled capillary with electrodes on either side, powered by a high voltage source that drives a large current through the lens. They have lengths of a few mm to many cm and aperture diameters a few hundred microns to mm-scale.

Naturally, the linearity of the focusing field is critical for preserving the beam emittance. When used with lighter gases such as hydrogen or helium, a temperature gradient between the center of the lens and the wall can cause a nonuniformity in the conductivity and thus the current density, leading to a nonlinear field and emittance increase~\cite{van_tilborg_nonuniform_2017,bobrova_simulations_2001,broks_nonlocal-thermal-equilibrium_2005,lindstrom_emittance_2018}. Nonlinearity in the field can also arise due to the $z$-pinch effect~\cite{bennett_magnetically_1934}, in which the drive current is self-focused towards the axis~\cite{autin_z-pinch_1987,christiansen_studies_1984,boggasch_z-pinch_1991}, reducing the radius over which the lens is linear.
This could present a limitation of the ultimate field strength in an active plasma lens.

On the other hand, it has been demonstrated that with a heavy gas like argon, APLs can have a uniform current distribution and thus a linearly increasing magnetic field, preserving the emittance~\cite{lindstrom_emittance_2018,lindstrom_erratum_2019}.
The linearity of the field has been demonstrated up to a gradient of 3.6~kT/m in a capillary with diameter of \SI{500}{\micro\meter}~\cite{sjobak_strong_2021}.

While the effective focusing strength of an APL is an order of magnitude higher than standard quadrupoles, it is ~2-3 orders of magnitude weaker than the UPL . However, APLs do not require a beam driver, making them compact and simple to operate~\cite{lindstrom_staging_2021}. Also, because the focusing effect arises entirely from the discharge current and not any self-driven wakefield effects (like the UPL), it is straightforward to use APLs to focus electrons or positrons. One only needs to switch the polarity of the current (i.e. put the high voltage electrode on either the upstream or downstream side of the APL).

A challenge in using APLs in focusing of high brightness, high intensity beams arises from the impact of self driven wakefields. In other words, passive plasma lensing will also occur in an APL if the beam density is sufficiently high. These generated fields are a complex function of both APL and beam parameters and generally are neither linear or easily tunable~\cite{lindstrom_analytic_2018}.
The impact of self driven wakefields can be mitigated to an extent by using shorter lenses with higher gradients, and shortening the plasma density ramps is one way of reducing this effect~\cite{kim_witness_2021}. 
Additionally, self driven wakefields have a strong dependence on the bunch length, with shorter bunches being more favorable. This scaling is naturally conducive to emerging interest in the use of attosecond-scale bunches for future high luminosity colliders ~\cite{Yakimenko:2019prl}.

\section{\label{sec:limits}Physical Limitations and Effects on Beam Spot Size}
\subsection{Oide Effect}

In 1988, Oide described the effect of hard synchrotron photon emissions in a strongly focusing quadrupole magnet on the emittance of a beam~\cite{Oide1988}.  This emittance growth results in a minimum achievable final spot size, referred to as the Oide limit.  This is not a hard limit, as the magnitude of the effect due to hard synchrotron radiation depends on the strength and length of the focusing optic, as well as the incoming beam parameters. In the Oide model, the minimum achievable final spot size is given by:
\begin{equation}\label{eqn:oidelimit}
    \sigma_y^*=\left(\frac{7}{5}\right)^{1/2}\left[\frac{275}{3\sqrt{6\pi}}r_e \lambdabar_e F(\sqrt K L, \sqrt K l^*)\right]^{1/7}(\epsilon_{Ny})^{5/7}
\end{equation}
with classical electron radius $r_e$, Compton wavelength $\lambdabar_e$, normalized emittance $\epsilon_N$, focusing gradient $K$, lens thickness $L$, distance from the focus to the lens's front end $l^*$, and where $F(\sqrt K L,\sqrt K l^*)$ is a dimensionless function that generally increases with focusing strength of the focusing optic and decreases with the thickness of the focusing optic.  While Eqn.~\ref{eqn:oidelimit} gives the minimum spot size, the more general spot size due to hard synchrotron radiation can be calculated using
\begin{equation}\label{eqn:oidespotsize}
    \sigma_y^{*2}=\beta_y^*\epsilon_y+\frac{110}{3\sqrt{6\pi}}r_e\lambdabar_e\gamma^5F(\sqrt K L,\sqrt K l^*)\left(\frac{\epsilon_y}{\beta_y^*}\right)^{5/2}
\end{equation}
where Eqn.~\ref{eqn:oidelimit} is recovered if the betafunction at focus $\beta_y^*$ is chosen as
\begin{equation}\label{eqn:oidebeta}
    \beta_y^*=\left(\frac{275}{3\sqrt{6\pi}}r_e\lambdabar_eF(\sqrt K L,\sqrt K l^*)\right)^{2/7}\gamma(\epsilon_{Ny})^{3/7}
\end{equation}

On one end of the spectrum, a very thin and dense plasma lens operating in the underdense regime and focusing a large electron beam would produce a significant amount of hard synchrotron radiation and the Oide limit would be quite large.  At the other end of the spectrum, the Oide limit could be completely suppressed if the incoming beam is capable of being adiabatically focused by slowly increasing the plasma density such that the betafunction decreases linearly~\cite{chen:1990prl}.  True adiabatic focusing from a plasma source could be difficult to achieve in practice, so future exploration into plasma-based final focusing systems will need to achieve a balance with lowering the Oide limit and generating a reasonable plasma source.

\begin{figure}[tb]
\centering
\includegraphics[width=5in, keepaspectratio]{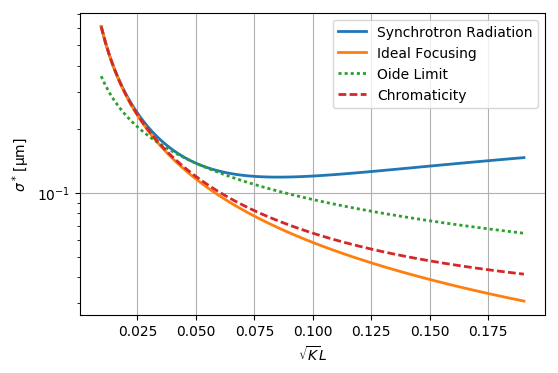}
\caption[]{Beam spot size at focus plotted against normalized thickness of the plasma lens.  Focusing strength $K$ is kept constant, and lens thickness $L$ varies from $10$ to $200\mathrm{\ \mu m}$.  Plotted in orange is the ideal spot size at the focus resulting from purely linear focusing with zero aberrations.  Red dashed represents the spot size if only the chromatic aberrations are considered.  Green dashed represents the Oide limit of Eqn.~\ref{eqn:oidelimit}, which decreases as the lens thickness is increased.  The blue curve is the spot size at the focus from Eqn.~\ref{eqn:oidespotsize}, and is what would be measured in a successful experiment to verify the Oide limit.}
\label{fig:oideplot}
\end{figure}

The Oide limit has yet to be reached experimentally, however an experiment could be designed in the near future to take advantage of the strong focusing gradient of an underdense plasma lens (Section ~\ref{sec:passivelens}) to experimentally verify Eqn.~\ref{eqn:oidespotsize}.  If we consider a $10\mathrm{\ GeV}$, two-bunch electron beam configuration, the Oide limit can be reached using an underdense plasma lens of density $n_p = 10^{18}\mathrm{\ cm^{-3}}$ and thickness in the range $L = 10-100$ $\mu$m..  This requires a high peak current drive bunch capable of driving a large blowout wake at such a high density, as well as a large witness bunch of $\beta_i = 500 \mathrm{\ cm}$ and $\epsilon_N=3\mathrm{\ \mu m-rad}$.  Additionally, the witness bunch requires a very low energy spread of $\delta_E=0.1\mathrm{\%}$ to minimize chromatic aberrations at such a large betafunction.  In Fig.~\ref{fig:oideplot} we plot the expected spot size due to aberrations from hard synchrotron radiation using these beam and plasma parameters.

There are some challenges and requirements necessary for such an experiment to be successful.  First, an underdense plasma lens must be generated with the necessary density, range of thicknesses, and be shown to focus according to theory under ideal circumstances.  Second, the required electron beam configuration needs to be generated at $10 \mathrm{\ GeV}$ and low energy spread.  The large witness bunch needs to fit within a very small wake at high plasma densities, and the drive bunch needs to be both high peak current and close to the witness beam longitudinally.  Third, a downstream imaging spectrometer needs to be capable of measuring beam spot sizes on the order of $100\mathrm{\ nm}$.  To ease some of these requirements, it may be possible to lower the plasma density and in return increase the beam's emittance for a larger spot size and increase the plasma lens thickness. On the other hand, the Oide limit is not a hard limit and there may be some benefit to ultra-aggressive focusing~\cite{Hirata1989}.

\subsection{Chromatic Effects}
\begin{figure}
    \centering
    \includegraphics[width=0.7\linewidth]{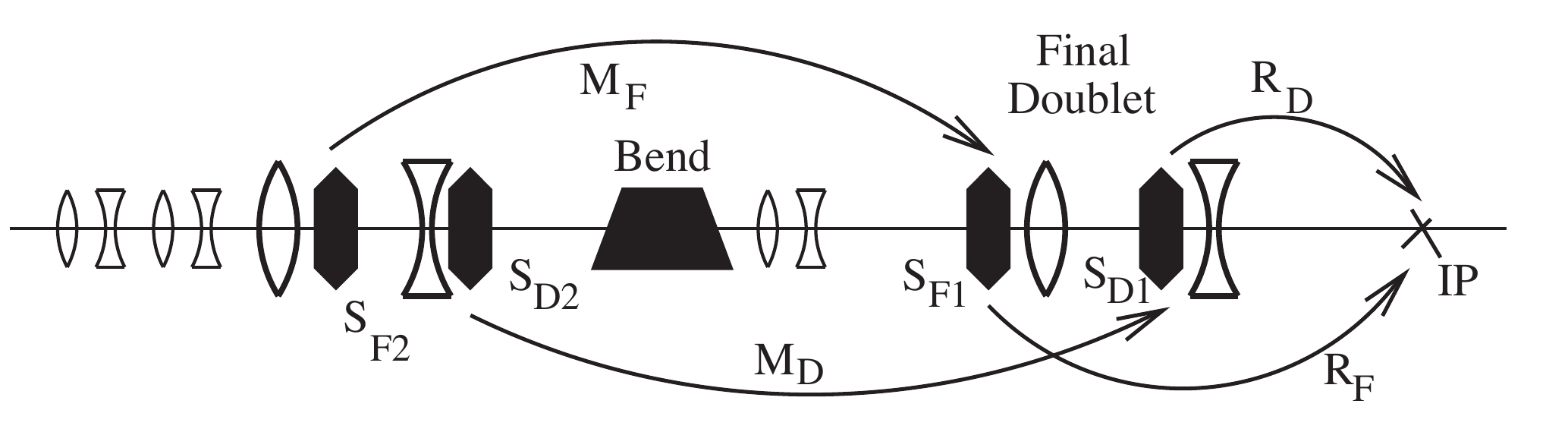}
    \caption{Chromaticity correction using a combination of sextupoles and dispersion from dipoles. From~\cite{Raimondi2001}.}
    \label{fig:chromaticity}
\end{figure}

In linear optics devices like quadrupoles or plasma lenses, particles of different energy have different focal lengths. This means that when a beam of finite energy spread is focused, not all particles reach the focal waist at the same time, which can severely limit the minimum spot size. The chromaticity of a focusing system is often quantified to first order using the chromatic amplitude
\begin{equation}
    W = \sqrt{\left(\frac{\partial\alpha}{\partial\delta}-\frac{\alpha}{\beta}\frac{\partial\beta}{\partial\delta}\right)^2 + \left(\frac{1}{\beta}\frac{\partial\beta}{\partial\delta}\right)^2},
\end{equation}
where $\alpha$ and $\beta$ are the Courant-Snyder parameters and $\delta$ is the relative energy offset from the nominal energy. The contribution from each focusing device is approximately $\Delta W = \beta/f$, where $f$ is its focal length, which means that in a beam-delivery system, the main source of chromaticity is the final doublet, where the beta function is large and the focal length is short. Final focusing systems are therefore designed specifically to compensate for this final large chromaticity with upstream optics. While it is possible to do so with linear optics alone, so-called apochromatic correction \cite{Montague1987,Lindstrom2016}, higher order chromatic effects often limit the energy bandwidth. Instead, the most potent solution is that employed by all modern final-focusing systems \cite{Raimondi2001}: nonlinear optics, specifically sextupoles in regions of large dispersion. Moreover, these sextupoles should be placed close to the strongest focusing elements, such as the final doublet, and must be paired with similar sextupoles upstream to cancel the detrimental effects of nonlinear focusing forces (see Fig.~\ref{fig:chromaticity}). This setup, known as local chromaticity correction, goes a long way in correcting the chromatic effects, but can be complex and sensitive to misalignment and strength errors. Therefore, reducing the chromaticity caused by the final doublet will always be beneficial. Plasma lenses are ideal in this regard, for two reasons: (1) The beam does not need to be defocused in one plane before the final focusing occurs, as it does in a quadrupole-based final doublet. The axisymmetric focusing of a plasma lens therefore suppresses the chromaticity in one of the two planes. (2) Plasma lenses are often physically small devices, which may allow them to be placed even closer to the interaction point. This could reduce the focal length $L*$ and therefore the chromaticity ($W \approx L*/\beta*$, since $\beta \approx L*/\beta*$ and $f = L*$). On the other hand, introducing plasma lenses close to the IP may cause emittance growth and a beam halo from scattering in the plasma.

\section{Beamstrahlung, GUINEA-PIG, and PIC Simulations}

\subsection{Beamstrahlung}

Beamstrahlung, radiation emitted during the beam-beam interaction, can limit the charge per bunch at the IP.  The beam-beam interaction is characterized by the Lorentz-invariant Beamstrahlung parameter \cite{Chao2012},
defined as the mean field strength in the beam rest frame $\gamma\left<E+B\right>$ in units of the Schwinger critical field $E_c=m^2c^3/e\hbar$. For a Gaussian beam, the average beamstrahlung parameter is
\begin{align}
    \Upsilon \approx \frac{5r_e^2}{6 \alpha} \frac{\gamma}{\sigma_z}  \frac{N}{(\sigma_x + \sigma_y)},
    \label{eq:Y}
\end{align}
where $\alpha$ is the fine structure constant.
For a given luminosity
\begin{align}
\mathcal{L}=\frac{fN^2}{4\pi\sigma_x\sigma_x}=\frac{P_b}{4\pi E_b}\frac{N}{\sigma_x\sigma_y},
\end{align}
fixed beam energy $E_b$ and beam power $P_b$, the beamstrahlung parameter Eq.~\eqref{eq:Y} is reduced by using flat beams, i.e., $\sigma_x/\sigma_y \gg 1$.  Beamstrahlung results in the emission of photons, unwanted background in the detector, and an increase in the effective beam energy spread owing to the energy loss.  
The number of photons emitted per lepton and the average relative energy spread induced by the beam-beam interaction in terms of the beamstrahlung parameter may be expressed \cite{Chao2012} as $n_\gamma \simeq 2.54\left(\alpha^2\sigma_z/r_e\gamma\right)\Upsilon\left(1+\Upsilon^{2/3}\right)^{-1/2}$ and $\delta_\gamma\simeq 1.24\left(\alpha^2\sigma_z/r_e\gamma\right)\Upsilon^2\left[1+\left(3\Upsilon/2\right)^{2/3}\right]^{-2}$, respectively.

Plasma and other advanced accelerator concepts propose to operate with ultrashort bunches.  For example, plasma accelerators naturally produce beams with lengths that are a fraction of the plasma wavelength, $\sigma_z < \lambda_p$ (with, e.g., $\lambda_p \sim 100~\mum$ for plasma densities $n \sim 10^{17}~{\rm cm}^{-3}$).  While conventional linear collider designs typically operate in the classical limit $\Upsilon \ll 1$, next-generation linear colliders using advanced accelerator technology with $\sqrt{s} > 1$~TeV will operate in the quantum beamstrahlung regime $\Upsilon \gg 1$, owing to the high beam energy and short bunch lengths. 
In this limit, $n_\gamma \propto \sigma_z \Upsilon^{2/3}$ and $\delta_\gamma\propto\sigma_z\Upsilon^{2/3}$.
Hence, assuming a fixed luminosity $\mathcal{L}$, center of mass energy $\sqrt{s}$, and fixed beam transverse sizes at the IP $\sigma_x$ and $\sigma_y$,
the number of photons and induced energy spread scale as $n_\gamma \propto N^{2/3}\sigma_z^{1/3}$ and $\delta_\gamma \propto N^{2/3}\sigma_z^{1/3}$.
On the other hand, for a fixed $\mathcal{L}$, $P_b$, $\sqrt{s}$ and transverse size ratio $\sigma_x/\sigma_y$, the number of photons and induced energy spread scale as $n_\gamma \propto \left(\sigma_z N\right)^{1/3}$ and $\delta_\gamma \propto \left(\sigma_z N\right)^{1/3}$.
Shorter beam lengths will reduce the beamstrahlung degrading effects at the IP.  Hence using advanced accelerator technology should allow one to operate at higher charge per bunch, reducing the power requirements to reach a target luminosity. 

\subsection{Trade-offs between bunch charge and repetition rate}

\begin{figure}[h!]
    \centering
    \includegraphics[width=0.7\linewidth]{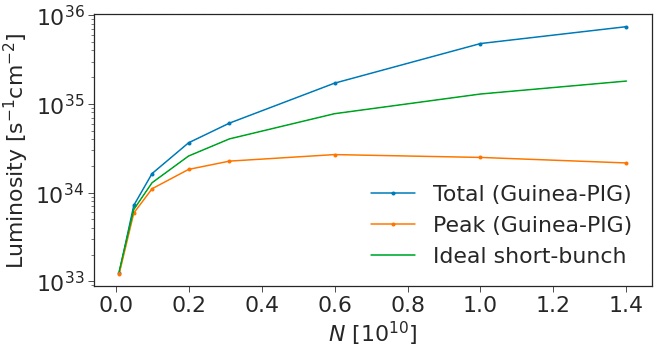}
    \caption{Calculated luminosity as a function of bunch charge. Total charge per second is kept constant by adjusting the repetition rate accordingly, i.e., $\frep\nb N=const$.}
    \label{fig:BunchChargeRepRateTradeOffs}
\end{figure}

A numerical simulation has been carried out at the interaction point to investigate possible trade-offs between the charge of individual bunches and the repetition rate of a linear collider. The luminosity calculation was performed using the Guinea-PIG \cite{schulte1997study} code, taking into account the pinch effect, beamstrahlung, pair creation. The results are presented in Fig.~\ref{fig:BunchChargeRepRateTradeOffs}.

In this simulation, the number of electrons (positrons) $N$ in one bunch is varied in the range from \SI{1.0e8}{} to \SI{1.4e10}{}. However, the total beam power is kept constant. This is achieved by keeping the total delivered charge per second constant (i.e., $\frep\nb N=const$) by changing the repetition rate $\frep$ accordingly. The number of bunches per train $\nb=\SI{208}{}$ is kept the same. The normalized transverse beam emittances and the bunch length were $\epsilon_x = \SI{0.66e-6}{m\,rad}$, $\epsilon_y = \SI{0.02e-6}{m\,rad}$, and $\sigma_z=\SI{44.0}{\micro m}$, respectively, at all values of $N$. The longitudinal density distribution was normal. The transverse beta functions at the interaction point were $\beta_x = \SI{7.1}{mm}$ and $\beta_y = \SI{0.15}{mm}$. Flat beam-beam collision was considered, $\sigma_x/\sigma_y\approx40$. The center of mass energy was $s_o=\SI{3.0}{TeV}$.

Figure~\ref{fig:BunchChargeRepRateTradeOffs} presents the total and the peak luminosity (blue and orange lines, respectively) calculated using Guinea-PIG. The peak luminosity is the luminosity calculated for the particles colliding with the energies above \SI{99}{\percent} of initial energy, i.e.,  $\sqrt{s}>0.99\sqrt{s_o}$. The green line corresponds to the ideal short-bunch approximation for the luminosity \cite{raubenheimer2000final}
\begin{equation}\label{eq:L00}
    \mathcal{L}_{00} = \frac{\frep \nb N^2}{4\pi \sigma_x \sigma_y},
\end{equation}
\noindent which does not account for luminosity enhancement, beamstrahlung, etc.

Figure~\ref{fig:BunchChargeRepRateTradeOffs} illustrates that the total luminosity and the ideal short-bunch luminosity monotonically increase with the number of particles in one bunch $N$, as expected. However, the peak luminosity has a maximum at a certain $N$. Indeed, after some $N$, the increase due to the $N^2$ dependence in Eq.~\eqref{eq:L00} is overpowered by the reduction due to beamstrahlung. This observation is important if the objective is to achieve higher peak luminosity. In the present SWFA design, $N=\SI{3.12e9}{}$ and $\frep=\SI{100}{Hz}$. Figure~\ref{fig:BunchChargeRepRateTradeOffs} shows that the maximum peak luminosity is achieved at $N=\SI{6e9}{}$ and $\frep\approx\SI{50}{Hz}$.

The presented simulation did not account for the likely change in bunch dimensions due to the change in bunch charge. Also, the bunches were assumed to be Gaussian, which is not exactly true due to the Oide effect \cite{oide1988synchrotron}. 

\subsection{\label{sec:extensions}Effects of Beam Aspect Ratio on Beam-beam effects}

   From the collider point of view, beams with a large aspect ratio (i.e. $\sigma_x/\sigma_y \gg 1$), or flat beams, have the advantage of reduced beamstrahlung effect at the interaction point, without sacrificing the peak luminosity. In a plasma accelerator, it is relatively unclear if flat beams will perform as well as round beams due to the asymmetry in beam loading, which could potentially lead to deterioration in beam quality. This is especially relevant for positrons, as no reliable acceleration regime has yet been identified. Since flat beams are considered to be supreme in reaching high peak luminosities in colliders, the effect of beam loading in accelerators for flat beams will be studied in parallel to achieve synergy. 
   
\begin{figure}
    \centering
    \includegraphics[width=0.4\linewidth]{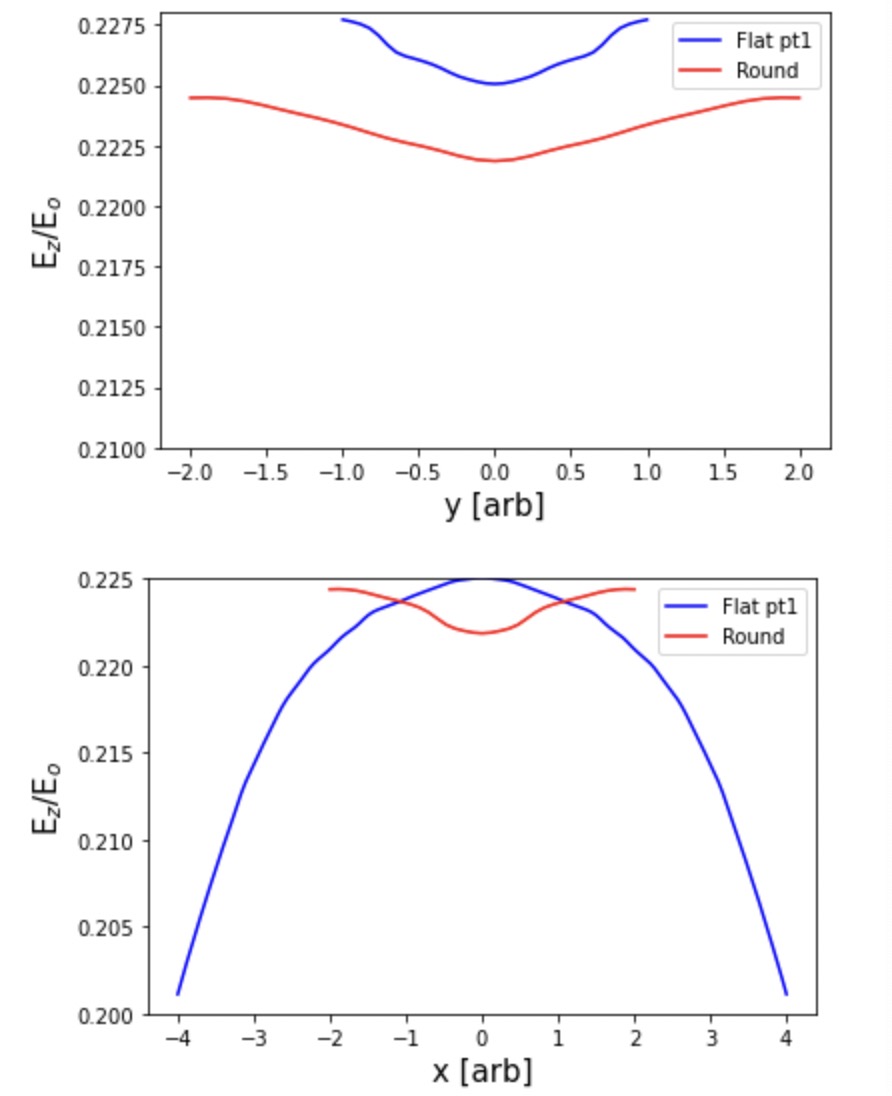} \includegraphics[width=0.4\linewidth]{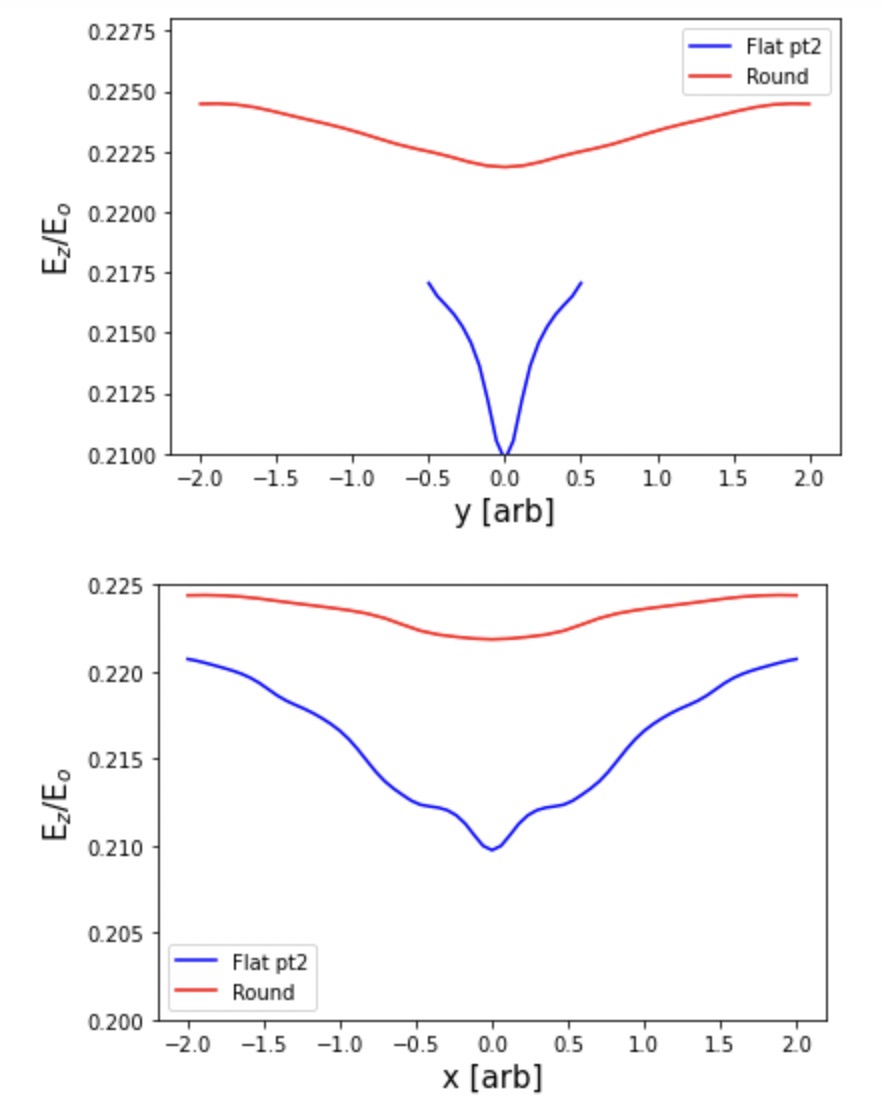}
    \caption{Left:Beams with $\sigma_x\sigma_y$=constant, thus keeping $\rho_b$ constant. The uncorrelated energy spread is a factor of 2.5 higher for a flat beam than for a round beam. Right: Beams with $\sigma_x$=constant and $\sigma_x/\sigma_y = \rho_{flat}/\rho_{round}$ for the flat beam. The uncorrelated energy spread is 11\% higher for a flat beam than for a round beam.}
    \label{fig:QPICAspRat}
\end{figure}

   A preliminary study on positron beam loading in a moderately non-linear regime for a round/flat beam is shown in FIG.\ref{fig:QPICAspRat}. The results demonstrate that for a flat beam with $\sigma_{x,flat} = \sigma_{x,round}$ and $\sigma_x/\sigma_y = \rho_{flat}/\rho_{round}$, the beam quality can be preserved to approximately the same level as the round beam. Here, only the uncorrelated energy spread is examined in accordance with \cite{PhysRevResearch.3.043063}. A more complete study on overall beam quality and energy efficiency will be performed. From the preliminary study, it is apparent that the beam density, in addition to the aspect ratio, could set an important limit on beam quality. This could in turn affect the possible flatness of the beam at IP. To be able to accommodate eventual positron acceleration schemes, both round and flat beams will be studied. Therefore, future work on asymmetrical beam collisions (i.e. colliding a flat electron beam with a round positron beam) is planned. 
    
\begin{figure}
    \centering
        \includegraphics[width=0.6\linewidth]{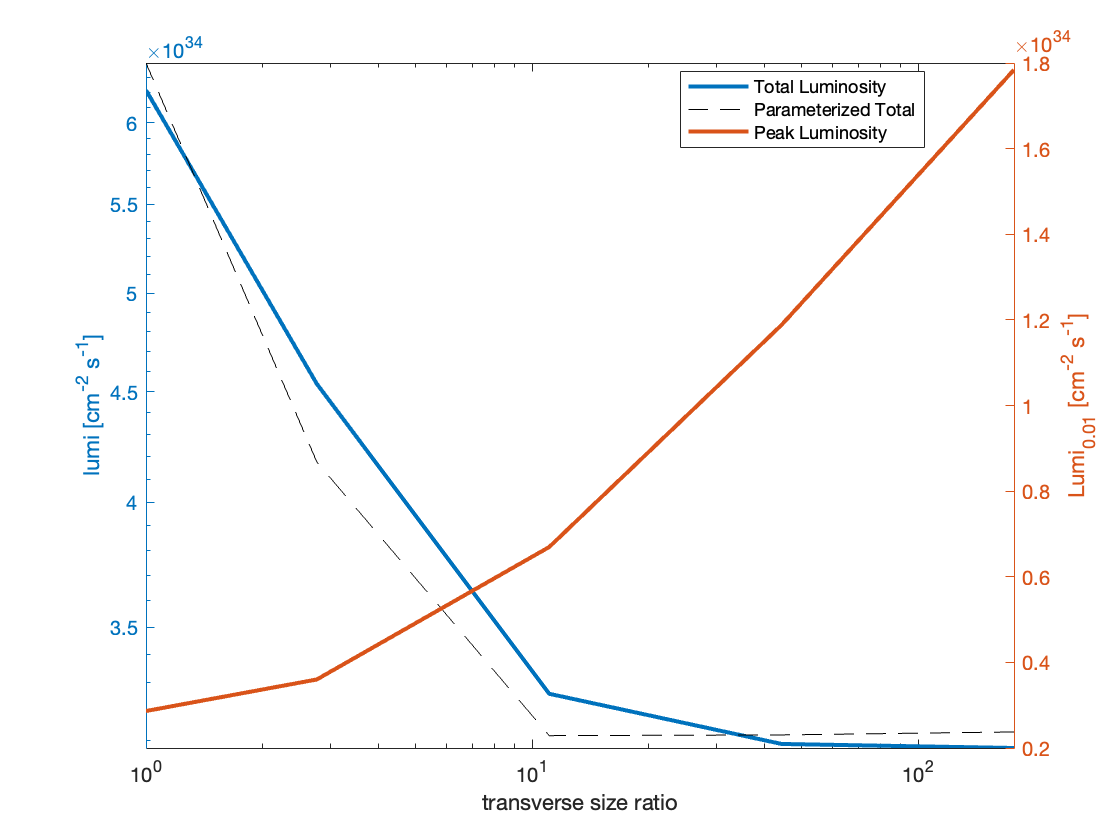}
        \caption{The simulated total and peak (i.e. the luminosity corresponding to particles colliding at $\sqrt{s}>0.99\sqrt{s_o}$) luminosity  over a range of aspect ratios ($\sigma_x\sigma_y$=constant). The parameterised total luminosity is calculated using the parameterization of the effective transverse sizes given in \cite{Chen:1992ax}.}
        \label{fig:AspRat}
    \end{figure}

     On the other hand, the quantitative effects on luminosity with varying aspect ratios need to be studied in detail and complement the flat/round beam study in a plasma accelerator. Previous studies parameterised the effect of aspect ratio in terms of the effective transverse sizes \cite{Chen:1992ax}. A preliminary comparison in luminosity performance for flat and round beams is made and the results are shown in FIG. \ref{fig:AspRat}. While round beams lead to higher total luminosity due to larger enhancement from mutual pinching between the two dimensions, the spread in the colliding beam energy spectrum due to beamstrahlung results in a much reduced peak luminosity compared to flat beams. The study aims to synergize the collider and accelerator sections for the beam in terms of the aspect ratio and find feasible parameter ranges that would satisfy both the luminosity and beam quality requirements.
    
\subsection{\label{sec:beamsAndQED}Non-linear QED and laser-free $\gamma-\gamma$ colliders}

In the large beamstrahlung regime with $\Upsilon > 1$, we observe significant dilution of the electron-positron luminosity spectrum from radiative losses. This is especially true in the round-beam regime. However, the large radiative losses also provide opportunities. First, they allow for the direct study of non-linear QED effects~\cite{DelGaudio2019}. Second, this opens the possibility of a laser-free $\gamma-\gamma$ collider where the colliding photons are generated from the beamstrahlung interaction~\cite{Yakimenko:2019prl,Tamburini2021,Zhang23}. Such a machine could provide the same physics as a $\gamma-\gamma$ collider with a much simpler design.

\subsection{Particle-in-Cell Simulations}

The state-of-the-art simulation codes for beam-beam collisions are the GUINEA-PIG~\cite{schulte1997study} and CAIN~\cite{Chen:1994jt} codes which have been in existence since the 1990s and 1980s, respectively. 
However, these codes are not well-suited to study high-disruption regimes, where QED effects produce vast amounts of electron-positron pairs, which eventually form a plasma around the collision point.
In addition, these codes may become more difficult to use and maintain over time, as their authors and users leave the field.

State-of-the-art Particle-in-cell~\cite{BirdsallBoolk2004, ArberPPCF2015, GonoskovPRE2015} (PIC) codes offer a new opportunity to use modern computing techniques to attack these decade old problems. The goal of beam-beam modeling in PIC codes is to provide descriptions of beam-beam interactions at the multi-TeV scale, but they will also provide a new tool for bench-marking and re-examining beam-beam parameters for traditional machines like the ILC.
As shown in Table~\ref{tab:features}, some of the major PIC codes in the community (e.g. Osiris~\cite{Osiris}, WarpX~\cite{MyersParallelComp2021}) already capture some of the QED effects that are required to study beam-beam collisions, and more of these are being currently implemented. In OSIRIS, the existing infrastructure allows for arbitrary QED processes driven by the beams or the lasers, namely pair production, beamstrahlung, Compton scattering and other processes already in place. 

In addition, we note that major PIC codes, such as Osiris and WarpX, are currently supported by large developer teams, and are constantly being updated to take advantage of the latest parallel computing architecture. An example is the WarpX code, part of the U.S. DOE Exascale Computing Project, which has been shown to scale on some of the world's largest CPU and GPU supercomputers\cite{FedeliGB2022}. OSIRIS has been run on many of the DOE and NSF leadership-class computing facilities for over 15 years.
It has scaled to the full machine on numerous previous systems including the Sequoia BlueGene machine at LLNL and Blue Waters where it achieved 2.2 Pflops on a full machine using > 10 trillion particles. Being able to leverage these high-performance computational resources for beam-beam collision studies would be advantageous.

Finally, full electromagnetic PIC codes are naturally better-suited to study the above mentioned high-disruption regimes. However, we note that even with PIC codes, these regimes can be computationally challenging due to the large number of macroparticles generated during the simulation. Computational strategies to reduce the number of macroparticles \cite{MuravievCPC2021,vranicCPC2015} are needed in this case, such as macroparticle culling or macroparticle merging.

As part of an effort to explore the use of modern PIC codes for beam-beam collisions, the WarpX team started benchmarking WarpX against GUINEA-PIG in selected collision scenarios. Furthermore, additional QED processes such as (linear) Breit-Wheeler, as well as macroparticle merging strategies, are being implemented in the code. 
The OSIRIS consortium has studied, using 3D PIC simulations, beamstrahlung radiation, and coherent pair creation in the mild quantum regime \cite{DelGaudio2019} and in the deep quantum regime \cite{Yakimenko:2019prl,Zhang23}. The numerical results agree with the theoretical predictions based on the strong-field QED theory framework. Moreover, the beamstrahlung photon spectrum obtained with OSIRIS has been benchmarked against GUINEA-PIG simulations in the mild quantum regime for low disruption, showing excellent agreement. The QED-PIC framework is a timely robust tool to perform self-consistent simulations of dense and relativistic beam-beam collisions in all regimes and for high disruption, where the feedback of additionally created pairs can act onto the self-fields of the beams. 
\begin{table}[h!]
\centering
\begin{tabular}{|l | c c c c|} 
 \hline
  Physics Process & GUINEA-PIG & CAIN & WarpX & OSIRIS \\ 
 \hline
 Quantum Synch. & \checkmark & \checkmark & \checkmark & \checkmark \\ 
 Bethe-Heitler & \checkmark & \checkmark & $\times$ & \checkmark \\
Linear Breit-Wheeler & \checkmark & \checkmark & $\times$ & $\times$ \\
Landau-Lifshitz & \checkmark & \checkmark & $\times$ &  $\times$ \\
Coherent Pair Production & \checkmark & \checkmark & \checkmark & \checkmark \\
Trident Cascade & \checkmark & $\times$ & $\times$ & $\times$ \\
Hadronic Production + Minijets & \checkmark & $\times$ & $\times$ & $\times$ \\
Electron-Laser Interaction & $\times$ & \checkmark & \checkmark & \checkmark \\
%$\gamma-\gamma$ Collisions & $\checkmark$ & \checkmark & $\times$ & $\times$ \\
 \hline
\end{tabular}
\caption{Features of traditional beam-beam codes (GUINEA-PIG, CAIN) and first-principles electromagnetic PIC codes (WarpX, OSIRIS). Physics processes that are currently missing from WarpX and OSIRIS are in the process of being implemented.}
\label{tab:features}
\end{table}

\section{\label{sec:MDI}Machine-Detector Interface}

The design of the beam delivery system and machine detector interface is an iterative process. An important starting point is a baseline estimate of radiation dosage as a function of distance from the IP. This study will aid the design of the inner detector layers, which in turn constrain the reconstruction of certain event signatures. 

Various elements of particle detectors can impact the design of the these focusing schemes. Detectors are composed primarily of tracking detectors to measure charged particle trajectories, calorimeters to measure energy deposits from charged and neutral particles, and muon systems designed to tag or measure muons which pass through the other systems. Additionally, an optimal design generally requires as close to $4\pi$ coverage of the solid angle as reasonably possible in order to fully measure particles produced in the collision. Elements of the focusing system primarily impact the tracking system and forward calorimetry.

Tracking detectors are composed of several subsystems: typically a silicon-pixel vertexing detector placed as close as possible to measure the precise decay vertices of displaced particles such as $B$-hadrons (which are critical to the study of Higgs bosons and top quarks), a gaseous or silicon-strip outer tracker to perform precise momentum measurements, and a solenoid to provide a uniform magnetic field to allow for momentum measurements. Typically, pixel vertexing detectors are placed as close as possible to the interaction point, to provide as best a measurement of the impact parameter of tracks as possible. For example, the Belle II pixel detector is placed at 14 mm~\cite{Belle-II:2010dht}, the ATLAS Insertable B-Layer is placed at 33 mm~\cite{Capeans:1291633}, and the CMS Phase-2 Upgrade is placed at 30 mm~\cite{CERN-LHCC-2017-009}. Bringing this layer as close as possible to the interaction point increases the radiation exposure of the detector, which can lead to limited lifetime or complete inoperability. Current designs for the HL-LHC are expected to be able to withstand 2000 fb$^-1$ of accumulated data, corresponding to 10 MGy and $2\times 10^{16} \mathrm{n}_\mathrm{eq}/\mathrm{cm}^2$ before replacement~\cite{CERN-LHCC-2017-021}. Depending on the precise nature of the beam-backgrounds induced by a plasma focus, the pixel vertexing detector may need to be placed at a further distance, worsening impact parameter resolution and degrading the identification of $B$-hadrons in data. Many of the most important measurements of the Higgs boson study the dominant $b\bar{b}$ decay channel, and therefore utilize $B$-hadron identification (via $b$-tagging) to reduce backgrounds. Reducing the performance of $b$-tagging, or significantly degrading it by removing pixel-vertexing detectors all together, could therefore have a substantially negative performance impact on important physics measurements. Experiments at hadron colliders, where radiation doses are significantly larger than the environments at electron-positron colliders, often place their vertexing systems at twice the radial distance.  This leads to degradation in performance, but identification of Higgs bosons decaying to $b$-jets has still been possible, even in the face of higher background rates~\cite{ATLAS:2018kot,CMS:2018nsn}: if simply changing the radial position of the detector would alleviate the worst of radiation damage from the plasma focusing device, then the physics goals of the machine still seem plausibly within reach.

The outer tracker is typically exposed to far less radiation because of its distance to the interaction point. Gaseous detectors such as Time Projection Chambers are extremely robust to radiation damage, and should be able to perform precise measurements of particle trajectories and momenta regardless of the radiation environment. If radiation is low enough, silicon-strip detectors can also be considered. Another possible area of concern, depending on the nature of focused-induced backgrounds, is the solenoidal magnetic field. These magnets are typically quite strong (2 T, or higher) and are therefore potentially sensitive to quenching from unexpected particle interactions. This can be mitigated by placing the magnet at a very large radius, but comes at a higher potential construction cost and impacts on calorimeter and flux return designs.

Finally, the physical focusing system is placed close to the beam pipe, away from the interaction point; this is typically where forward calorimeters are placed to increase the solid-angle coverage of a detector. This coverage is important for many crucial measurements, such as the exploitation of the full hadronic recoil to measure inclusive Higgs production~\cite{Thomson:2015jda}. Reducing this detector coverage may degrade the resolution of these measurements, requiring greater integrated luminosity to achieve the same physics targets as other colliders.

Concluding, the radiation environment produced by the final focus system can provide substantial challenges to the design of detectors, but many strategies exist to mitigate the effects and extract useful physics measurements. 

\section{Conclusion and Continuing Research}

% Simulation resources for nanometer size beams

The Snowmass'21 process brought together physicists from the Advanced Accelerator and High Energy Physics communities to discuss goals and challenges for the field. These discussions exposed the need for a dedicated study of Beam Delivery Systems and beamstrahlung effects, as these are critical areas of research for multi-TeV $e^+e^-$ and $\gamma-\gamma$ colliders. In addition, the Advanced Accelerator community must work closely with High Energy theorists and experimentalists to advance the design of these systems based on physics benchmarks.

We have formed a collaboration to focus on PIC simulations for beamstrahlung physics and optimization studies. The goals of this collaboration include:
\begin{enumerate}
    %\item Inclusion of all relevant QED processes.
    %\item Modeling of Compton scattering for $\gamma-\gamma$ with relevant non-linear effects.
    \item Comparisons with ILC benchmarks at 500 GeV CM.
    \item Studies at large disruption parameter for ILC-type beams.
    \item Comparisons with CLIC benchmarks at 3 TeV CM.
    \item Extrapolation to the multi-TeV regime.
    \item Beamstrahlung mitigation studies including:
    \begin{itemize}
        \item Beam aspect ratio studies.
        \item Bunch length studies.
        \item Neutral beam studies.
        \item Modulated beam studies.
    \end{itemize}
\end{enumerate}
The first goal, comparing with ILC benchmarks, will demonstrate the applicability of PIC codes for beam-beam collisions and provide a valuable new tool to the HEP community. This is a crucial first step to the collaboration's long-term goal of simulating collisions at 15 TeV CM, where accurate predictions of the luminosity spectrum will be needed to estimate the power consumption of the machine.

Finally, new experimental programs at FACET-II aim to demonstrate the viability of passive plasma lenses~\cite{Doss2019} and explore non-linear QED effects in beam-laser interactions~\cite{Chen2022}. The latter experiment is a crucial first step toward a Compton IP for a $\gamma-\gamma$ collider. These new experimental and simulation efforts will improve our understanding of BDS systems for multi-TeV linear colliders and aid in the design and costing of the system for future machines.

\acknowledgments

This work was supported by the Research Council of Norway (NFR Grant No. 313770 and 310713). This work was supported by the Director, Office of Science, Office of High Energy Physics, of the U.S. Department of Energy under Contract No. DE-AC02-05CH11231. Work supported by the U.S. Department of Energy under Contract DE-
AC02-76SF0051.

% Bibliography
%% [A] Using JHEP.bst file
\bibliographystyle{JHEP}
\bibliography{main.bib}

%% or
%% [B] Manual formatting (see below)
%% (i) We suggest to always provide author, title and journal data or doi:
%% in short all the informations that clearly identify a document.
%% (ii) please avoid comments such as "For a review'', "For some examples",
%% "and references therein" or move them in the text. In general, please leave only references in the bibliography and move all
%% accessory text in footnotes.
%% (iii) Also, please have only one work for each \bibitem.

%%\begin{thebibliography}{99}

%%\bibitem{a}
%%Author,
%%\emph{Title},
%%\emph{J. Abbrev.} {\bf vol} (year) pg.

%%\bibitem{b}
%%Author,
%%\emph{Title},
%%arxiv:1234.5678.

%%\bibitem{c}
%%Author,
%%\emph{Title},
%%Publisher (year).

%%\end{thebibliography}
\end{document}